\documentclass[a4paper]{article}
\usepackage{a4wide}
\usepackage{authblk}
\usepackage{setspace}
\usepackage{lineno}
\usepackage{mathrsfs,amsmath,amssymb,ascmac,cases,bm}
\usepackage{mathtools}
\usepackage{braket}
\usepackage{chngcntr}
\usepackage{bm}% bold math

\newcommand{\pd}[2]{\frac{\partial #1}{\partial #2}}

\newcommand{\spd}[2]{\partial #1/\partial #2}

\newcommand{\inv}[1]{{#1}^{-1}}

\newcommand{\Pare}[1]{\left(#1 \right)}

\newcommand{\Angle}[1]{\left\langle#1 \right\rangle}

\newcommand{\R}[1]{\mathbb{R}^{#1}}
\newcommand{\df}[2]{\Omega^{#1}(#2)}
\newcommand{\sect}[1]{\Gamma\left({#1}\right)}
\newcommand{\bms}[1]{#1_{\#}}
\newcommand{\embr}{x}
\newcommand{\embc}{y}
\newcommand{\bi}{\vartheta}
\newcommand{\rb}[1]{E_{#1}}
\newcommand{\unitnormal}[1]{\mathcal{N}^{#1}}

\newcommand{\M}{\mathcal{M}}

\newcommand{\metric}[1]{g[#1]}
\newcommand{\ac}[1]{\nabla[#1]}

\newcommand{\bsb}[3]{B_{\Pare{{#1}, {#2}, {#3}}}} % Bs-pline
\makeatletter
\newcommand{\setlabel}[2]{\def\@currentlabel{#2}\label{#1}}
\makeatother
\begin{document}

%%%% Article title to be placed here
\title{
Geometrical frustration in nonlinear mechanics of screw dislocation
}

\author[1]{Shunsuke Kobayashi}
\author[1]{Ryuichi Tarumi}

\affil[1]{Graduate School of Engineering Science, Osaka University, 1-3 Machikaneyama-cho, Toyonaka, Osaka, 560-8531, Japan}
\date{}
\maketitle

%%%% Abstract text to be placed here %%%%%%%%%%%%
\begin{abstract}
The existence of stress singularities and reliance on linear approximations pose significant challenges in comprehending the stress field generation mechanism around dislocations.
This study employs differential geometry and calculus of variations to mathematically model and numerically analyse screw dislocations.
The kinematics of the dislocation are expressed by the diffeomorphism of the Riemann--Cartan manifold, which includes both the Riemannian metric and affine connection.
The modelling begins with a continuous distribution of dislocation density, which is transformed into torsion $\tau$ through the Hodge duality.
The plasticity functional is constructed by applying the Helmholtz decomposition to bundle isomorphism, which is equivalent to the Cartan first structure equation for the intermediate configuration $\mathcal{B}$.
The current configuration is derived by the elastic embedding of $\mathcal{B}$ into the standard Euclidean space $\mathbb{R}^3$.
The numerical analysis reveals the elastic stress fields effectively eliminate the singularity along the dislocation line and exhibit excellent conformity with Volterra's theory beyond the dislocation core.
Geometrical frustration is the direct source of dislocation stress fields, as demonstrated through the multiplicative decomposition of deformation gradients.
By leveraging the mathematical properties of the Riemann--Cartan manifold, we demonstrate that the Ricci curvature determines the symmetry of stress fields.
These results substantiate a long-standing mathematical hypothesis: the duality between stress and curvature.
\end{abstract}

\maketitle
\noindent

\section{Introduction}

Plastic deformation in crystalline materials primarily occurs via slip deformation among close-packed crystal planes caused by displacements of one part of a crystal relative to another part along these planes.
The non-uniform slip deformation along the sliding direction leads to the formation of a line defect, referred to as a dislocation.
Dislocations are critical in determining the mechanical properties of crystalline materials, including strength, ductility, creep resistance, and fracture toughness \cite{Hirth}.
Early studies \cite{orowan3,polanyi,taylor,Volterra} significantly contributed to comprehending dislocation mechanics.
However, despite providing valuable analytical expressions for dislocation stress fields \cite{Volterra, Mura}, several challenges remain to be addressed.
One significant problem is the existence of stress singularities, which are regions where the stress components diverge to infinity within the dislocation core \cite{Mura}.
Equally important is the reliance on linear approximations.
Recently, extensive studies have addressed singularity in linearised systems \cite{lazar_1,lazar_2,lazar_3,lazar_4}; however, nonlinear finite deformation occurs in the proximity of the dislocation core.
Linear elasticity is valid only for infinitesimal deformations that preserve the material frame indifference \cite{Marsden-Hughes}.
Hence, the fundamental question of the stress field generation mechanism around dislocations remains unresolved.

Kondo's analysis of compatibility conditions in the early 1950s initiated a paradigm shift in the geometric modelling of dislocations.
Kondo highlighted that the degree of \textit{incompatibility} corresponds to the Riemann curvature \cite{kondo_1}.
Subsequently, he proposed a theory of dislocations based on differential geometry \cite{kondo_2}.
An equivalent geometrical theory of dislocations was independently proposed by Bilby \textit{et al.} \cite{bilby_continuous_1956,bilby_notitle_1955}, and Kr\"oner and Seeger \cite{kroner_nicbt-lineare_nodate}.
The mathematical equivalence of these theories was later established by Amari \cite{amari}.
In ordinary Euclidean space $\mathbb{R}^3$, a continuum cannot exist in an incompatible state. However, in a mathematically generalised manifold, such a state is possible.
These studies established the correlation between lattice defects, specifically dislocations and disclinations, and torsion and curvature in the affine connection\cite{kondo_2, bilby_continuous_1956,bilby_notitle_1955, kroner_nicbt-lineare_nodate, amari, anthony_theorie_1970}.
Noll \cite{noll}, Wang \cite{wang}, de Wit \cite{dewit}, Le and Stumpf \cite{le_stumpf_1,le_stumpf_2}, Wenzelburger \cite{wenzelburger}, and Binz \textit{et al.} \cite{binz} further developed the theory.
Recently, Yavari and Goriely reformulated the theories using modern differential geometry on the Riemann--Cartan manifold \cite{yavari_riemanncartan_2012,yavari_riemanncartan_2013,yavari_geometry_2014}.
Edelen \cite{edelen}, Acharya \cite{acharya}, and Clayton \cite{clayton} have performed similar studies,
albeit with significant restrictions on the dislocation arrangement to obtain an analytical solution.
Therefore, these methods are inadequate for determining stress fields for arbitrary dislocation configurations.
In subsequent sections, we will demonstrate the manner in which the geometrical theory of dislocation necessitates the solution of the Cartan first structure equation.
A standard solution to this problem is to reconstruct the Cartan equation in a variational form and solve it numerically instead of analytically.
This approach enables nonlinear stress field analysis for an arbitrary dislocation configuration, thereby broadening the scope of applications of the geometrical theory.
Moreover, the effect of free boundary on the mechanical fields of dislocation can be taken into consideration.
The approach can also provide a mathematical explanation for the stress field generated around dislocations, which has remained unresolved for a considerable period.

Considering this, we aim to develop a mathematical framework to determine the stress field of dislocations for arbitrary configurations.
By utilising the aforementioned novel theoretical framework based on differential geometry, we elucidate the mathematical basis of the stress field origins.
This study is structured as follows.
First, this introduction provides a succinct summary of the dislocation theory.
We have presented the current state-of-the-art and identified the outstanding challenges that require further attention.
In the following section, we will elucidate the kinematics of dislocation using differential geometry.
The Riemann--Cartan manifold provides a unified framework for expressing the three fundamental configurations: reference, intermediate, and current.
Section 3 provides variational formulations of the Cartan first structure and stress equilibrium equations.
Here, the Helmholtz decomposition is essential and hence, has been briefly discussed.
Section 4 presents the results of the numerical analysis for screw dislocation.
We initially present the distribution of the plastic deformation and the Riemannian metric for quantitative validation of our results.
We then present the nonlinear stress fields, which remain non-singular even in the dislocation core.
Section 5 discusses the mechanical origin of the dislocation stress field.
Here, we provide a comprehensive understanding of the geometrical frustration and internal stress field formation using the Ricci curvature.
In Section 6, we summarise our findings.

\section{Kinematics of dislocation on the Riemann--Cartan manifold}
\subsection{Reference and current configurations}

Following previous studies \cite{kondo_2, bilby_notitle_1955, kroner_nicbt-lineare_nodate, amari, yavari_riemanncartan_2012}, we apply differential geometry to describe the kinematics of dislocations.
Initially, we present three distinctive configurations: the reference $\mathcal{R}$, intermediate $\mathcal{B}$, and current $\mathcal{C}$ configurations.
They represent the three different states of a crystal: dislocation-free perfect crystal state, plastically deformed state due to dislocations, and elastically relaxed state following the plastic deformation.
While the reference and current configurations exist in the standard Euclidean space $\mathbb{R}^3$, the intermediate configuration cannot exist in $\mathbb{R}^3$ due to the presence of torsion within the dislocation core.
Thus, the Riemann--Cartan manifold \cite{yavari_riemanncartan_2012} provides a unified geometrical framework for describing kinematics.
Here, the key concept is to represent the three distinctive configurations using diffeomorphisms of a single manifold $\mathcal{M}$.

Let $\mathcal{M}$ be a three-dimensional compact $C^\infty$ manifold with a piece-wise smooth boundary, and let $g$ and $\nabla$ be the Riemannian metric and $g$-compatible affine connection.
Then, the triplet $(\mathcal{M}, g, \nabla)$ is referred to as Riemann--Cartan manifold.
The reference configuration $\mathcal{R}$ is defined by $C^\infty$-diffeomorphic embedding of the manifold into the Euclidean space such that $\embr{}: \M \to \mathbb{R}^3$.
This indicates that, for any $p\in \M$, the image $x(p)=(x^1, x^2, x^3)$ describes the Cartesian coordinate of the point in $\mathbb{R}^3$.
The map naturally endows the orthonormal basis $\spd{}{x^i}$ and its dual $dx^i$ that satisfy $dx^i(\spd{}{x^j})=\delta^i_j$.
The embedding map $\embr{}$ induces the Riemannian metric $g[x]$, which is locally represented by
\begin{align}\label{reference metric}
    \metric{x} = \delta_{ij} dx^i\otimes dx^j.
\end{align}
Let us express the trivial Euclidean connection for the reference configuration by $\nabla [x]$.
Subsequently, the Riemann-Cartan manifold for the reference configuration $\mathcal{R}$ is defined by the triplet ($(\M, \metric{\embr{}}, \ac{\embr{}})_{\mathcal{R}}$).
Because the connection $\nabla[x]$ has zero torsion $T=0$ and curvature $R=0$, this state is characterised as a Euclidean sub-manifold.

The current configuration $\mathcal{C}$ can be defined using a similar method.
Let $y: \M \to \mathbb{R}^3$ be the total deformation defined by the composition map of the plastic and elastic deformations.
Let $y$ be a map that induces the Riemannian metric $g[y]$ such that
\begin{align}\label{eq:CurrentMetric}
  \metric{y} = \delta_{ij}dy^i\otimes dy^j=\delta_{kl} F^k_i F^l_j dx^i\otimes dx^j,
\end{align}
where $F^i_j=\partial y^i/\partial x^j$ is the total deformation gradient.
The linear transformation of the dual basis $dx^i$ to the current configuration is defined by equation (\ref{eq:CurrentMetric}), such that $dy^i=F^i_jdx^j$.
Let $\nabla [y]$ be the Euclidean connection for the metric $g[y]$.
Then, the current configuration $\mathcal{C}$ is described by the triplet $(\mathcal{M}, g[y], \nabla[y])_\mathcal{C}$.
Again, this configuration is a Euclidean sub-manifold.

\subsection{Intermediate configuration}
Unlike the previous two cases, the construction of the intermediate configuration $\mathcal{B}$ necessitates the use of several mathematical properties of the Riemann--Cartan geometry.
The intermediate configuration denotes a state of plastic deformation without concurrent elastic deformation.
Consequently, the affine connection contains non-zero torsion \cite{yavari_geometry_2014,yavari_riemanncartan_2012, yavari_riemanncartan_2013}.
To initiate, we impose the fundamental mathematical assumption that the manifold $\mathcal{M}$ is \textit{parallelizable} \cite{wenzelburger}.
This ensures the existence of bundle isomorphism $\bi\colon T\mathcal{M}\to \mathcal{M} \times \mathbb{R}^3$, which is defined as $\mathbb{R}^3$-valued 1-form throughout $\mathcal{M}$.
In this study, the bundle isomorphism $\bi$ represents the plastic deformation gradient $F_p$ resulting from dislocation.
Based on the standard mathematical assumption, the Riemannian metric on the intermediate configuration can be introduced such that
\begin{align}\label{eq:IntermediateMetric}
  \metric{\bi} {}&=\delta_{ij}\bi^i\otimes \bi^j=\delta_{kl} \bi^k_i\bi^l_j dx^i\otimes dx^j,
\end{align}
where $\bi^i =\bi^i_j dx^j=(F_p)^i_j dx^j$ represents the linear transformation of the dual basis $dx^i$ from the reference to the intermediate configuration.
The Cartan moving frame on the intermediate configuration is represented by $(\vartheta^1, \vartheta^2, \vartheta^3)$.

Let $X, Y\in T\M$ be smooth vector fields defined on $\M$.
According to the standard theory of differential geometry, the bundle isomorphism $\bi$ induces an affine connection $\nabla[\bi]$ such that \cite{wenzelburger}
\begin{align}\label{eq:IntermediateConnection}
    \nabla[\bi]_X Y{}&= X^k \bigg(Y^j \Gamma^l_{jk}[\bi]+\pd{Y^j}{x^{k}}\bigg)
  \pd{}{x^{j}},
\end{align}
where $\Gamma^k_{ij}[\bi]$ represent the connection coefficients.
Two affine connections, Weitzenb\"ock connection $\nabla_W$ and Levi--Civita connection $\nabla_{LC}$, compatible with the Riemannian metric \cite{yavari_riemanncartan_2012} are introduced following the mathematical property of Riemann--Cartan manifolds.
This indicates that two distinct mathematical representations exist for the intermediate configuration: Weitzenb\"ock manifold $(\M, g[\bi], \nabla_W[\bi])_\mathcal{B}$ and Riemannian manifold $(\M, g[\bi], \nabla_{LC}[\bi])_\mathcal{B}$.
Coefficients of the two affine connections are obtained from the following forms \cite{tu_differential_2017}
\begin{align}\label{eq:ConnectionCoefficients}
\Gamma^l_{jk}[\bi] = \big(\inv{\bi}\big)^l_i \pd{\bi^i_j}{x^{k}},
\qquad
\Gamma^k_{ij}[\bi] = &{} \frac{g[\bi]^{kl}}{2}\left(
\pd{g[\bi]_{lj}}{x^i}+\pd{g[\bi]_{li}}{x^j} - \pd{g[\bi]_{ij}}{x^l}\right).
\end{align}
Notably, the coefficients $\Gamma^l_{jk}[\bi]$ obtained in equation (\ref{eq:ConnectionCoefficients})$_1$ yield the Weitzenb\"ock connection $\nabla_W[\bi]$, whereas the Levi-Civita connection $\nabla_{LC}[\bi]$ is obtained from equation (\ref{eq:ConnectionCoefficients})$_2$.
Because these manifolds share the Riemannian metric $g[\bi]$, their kinematical states are inherently indistinguishable.
The mathematical expressions of the incompatibility with Euclidean geometry are the critical difference.
The Weitzenb\"ock manifold employs torsion $T$, whereas the Riemannian manifold utilises curvature $R$.
Generally, the torsion is defined by $T(X, Y)\coloneqq \nabla_X Y-\nabla_YX - [X, Y]$, where $[\cdot, \cdot]$ denotes the Lie bracket.
Similarly, the curvature is expressed as $R(X, Y)Z
:=\nabla_X \nabla_Y Z-\nabla_Y \nabla_X Z-\nabla_{[X,Y]}Z$.
The local representations of $T$ and $R$ are given by \cite{nakahara_geometry_2003}
\begin{align}
\label{eq:torsion}
    T =&{}
      \Pare{\Gamma^i_{jk}-\Gamma^i_{kj}} \pd{}{x^i}\otimes dx^j\otimes dx^k,\\
\label{eq:curvature}
    R =&{} 
        \Pare{\pd{\Gamma^i_{jl}}{x^k} - \pd{\Gamma^i_{jk}}{x^l}
        + \Gamma^i_{mk}\Gamma^m_{jl}-\Gamma^i_{ml}\Gamma^m_{jk}}
        \pd{}{x^i}\otimes dx^j\otimes dx^k\otimes dx^l.
\end{align}
For future reference, we calculate the exterior derivative of the bundle isomorphism $\bi$ such that
\begin{align}\label{exderivative0}
  d\bi
  = \sum_{j<k} \left(\frac{\partial\bi^i_k}{\partial x^{j}}-\frac{\partial\bi^i_j}{\partial x^{k}}\right)dx^{j}\wedge dx^{k}\otimes \rb{i},
\end{align}
where $\rb{i}$ represents the orthonormal basis on $\R{3}$.
This external derivative $d\bi$ represents the torsion 2-form.
A direct computation demonstrates that the derivative $d\bi$ is equivalent to the torsion $T$ of the connection $\nabla[\bi]$.
Additional mathematical descriptions for the intermediate configuration are provided in \ref{AppendixA}\ref{sec:A1}.

\section{Governing equations and numerical analysis}
\subsection{Cartan first structure equation}

The dislocation density tensor $\alpha$ \cite{kondo_2, bilby_notitle_1955, kroner_nicbt-lineare_nodate} is the most fundamental quantity in the geometrical theory of dislocations.
To simplify the analysis, we consider a screw dislocation with the Burgers vector field $b=b^i \rb{i}$ and the tangent vector field of the dislocation line $n=n^j \delta_{jk}dx^k$.
Then, the dislocation density field $\alpha \in \Omega^1 (\M; \mathbb{R}^3)$ is expressed by the following form \cite{nye}
\begin{align}\label{eq:dislocation_density}
\alpha= f b^i n^j \delta_{jk} dx^k\otimes \rb{i}.
\end{align}
Here, $f$ denotes the two-dimensional distribution function around the dislocation core.
Notably, the superposition of the equation (\ref{eq:dislocation_density}) can be utilised to analyse multiple dislocations.
In the classical theory of dislocation, $f$ is commonly represented by the Dirac delta function \cite{Mura}.
However, the discontinuity of the delta function in this formulation results in stress singularities at the dislocation core.
To address this issue, substituting the delta function with a continuous function is imperative.
The operation introduces torsion within the dislocation core, which is incompatible with the standard Euclidean geometry.
This observation highlights the significance of differential geometry in dislocation analysis.

Previous studies demonstrated the equivalence between the dislocation density tensor $\alpha$ and the torsion 2-form $\tau \in \Omega^2 (\M; \mathbb{R}^3)$ \cite{bilby_continuous_1956}.
In modern notation, the relation is expressed as $\tau=*\alpha$, where $*$ denotes the Hodge star operator \cite{yavari_riemanncartan_2012}.
According to equation (\ref{eq:dislocation_density}), the distribution of torsion 2-form around the screw dislocation becomes
\begin{align}\label{eq:dislocation-torsion}
\tau=*\alpha=
\sum_{j<k}f b^i n^{l}\epsilon_{ljk} dx^j\wedge dx^k \otimes \rb{i},
\end{align}
where $\epsilon_{ljk}$ represents the permutation symbol.
The result is the torsion 2-form $d\bi$ of the intermediate configuration in equation (\ref{exderivative0}).
The celebrated Cartan first structure equation is derived from this profound insight:
\begin{align}\label{eq:cartan}
  \tau=d\bi.
\end{align}
This equation connects the kinematics of dislocation and the Riemann--Cartan geometry.
Because the left-hand side of equation (\ref{eq:cartan}) can be obtained from equation (\ref{eq:dislocation-torsion}), integration of the external derivative yields the bundle isomorphism $\bi$ (or plastic deformation gradient $F_p$).
Consequently, the intermediate configuration $(\M, g[\bi], \nabla [\bi])_\mathcal{B}$ that is responsible for plastic deformation due to the dislocation density $\alpha$ can be determined. 
Notably, the integration of $\bi$ along a closed curve $\partial S$ on $\M$ obtains the Burgers vector such that
\cite{wenzelburger}:
\begin{align}\label{eq:Burgers_circuit}
b[S]=\int_{\partial S}\vartheta^i \otimes \rb{i}=\int_{S}d\vartheta^i \otimes \rb{i}.
\end{align}
Here, we obtain the Bianchi identity $d\tau=0$ since $d\tau=dd\bi= 0$.
This outcome indicates that the dislocation does not terminate within the material.

\subsection{Helmholtz decomposition and plasticity functional}

Cartan's first structure equation (\ref{eq:cartan}) establishes the equivalence between torsion 2-forms derived from the bundle isomorphism and the dislocation density.
To enable analysis, we developed a novel numerical approach based on the calculus of variations.
Here, the Helmholtz decomposition is essential.
It guarantees that for all $\R{3}$-valued 1-forms $\df{1}{\M; \R{3}}$, we obtain the orthogonal decomposition such that \cite{wenzelburger, binz, schwarz}
\begin{align}
    \df{1}{\mathcal{M};\mathbb{R}^3} =dC^\infty(\mathcal{M};\mathbb{R}^3)\oplus \mathcal{D}(\mathcal{M};\mathbb{R}^3),
\end{align}
where $dC^\infty(\mathcal{M};\mathbb{R}^3)$ and $\mathcal{D}(\mathcal{M};\mathbb{R}^3)$ represent the $\mathbb{R}^3$-valued exact 1-forms and dual, respectively (see Appendix \ref{sec:HelmholtzDecomposition}).
As explained in a previous section, the bundle isomorphism $\bi$ has the $\mathbb{R}^3$-valued 1-form.
Hence, Helmholtz decomposition of the bundle isomorphism becomes
\begin{align}\label{eq:Helmholtz decomposition}
    \bi^{} = d\psi+\Theta = \left(\frac{\partial\psi^i}{\partial x^j}+\Theta^i_j\right)dx^j\otimes E_i,
\end{align}
where $\psi \in C^\infty(\mathcal{M};\mathbb{R}^3)$ and $\Theta \in \mathcal{D}(\mathcal{M};\mathbb{R}^3)$.
Consequently, the Cartan equation becomes
\begin{align}
    \tau=d\bi=d(d\psi+\Theta)=d\Theta,
\end{align}
since $d(d\psi)=0$ by definition.
This outcome indicates that only the dual exact part $\Theta$ is responsible for torsion $\tau$, whereas the exact part $\psi$ plays a less significant role.
Therefore, without loss of generality, we can set $\bi^i=dx^i+\Theta^i$, where $\psi=x$ denotes the identity map.

The Helmholtz decomposition also imposes a proper boundary condition for the dual exact part such that $\Theta |_{\partial \M}(\unitnormal{})=0$, where $\unitnormal{}$ denotes unit surface normal.
Consequently, we can introduce the plasticity function such that
\begin{align}\label{eq:plasticityfunctional0}
    I[\Theta] = 
    \int_\M\frac12 \Angle{\tau-d\Theta, \tau-d\Theta }
    +\int_\M  \Angle{\gamma, -\delta \Theta},
\end{align}
where $\gamma \in \df{0}{\M; \R{3}}$ is the Lagrange multiplier function.
Here, the first integrand represents the quadratic form of the residual of the Cartan first structure equation, and the second represents a constraint condition to ensure that $\Theta$ satisfies the requirement of being a dual exact form, i.e., $-\delta \Theta=0$ on $\M$.
Therefore, the solution $\Theta$ of the Cartan first structure equation (\ref{eq:cartan}) is defined to minimise the plasticity functional $\mathcal{I}[\Theta]$ as given by equation (\ref{eq:plasticityfunctional0}).
The present variational framework enables the construction of the intermediate configuration $\mathcal{B}$ for an arbitrary configuration of dislocations.

\subsection{Minimisation of strain energy functional}

The intermediate configuration $\mathcal{B}$ is distinguished by the plastic deformation, denoted by the bundle isomorphism $\bi$.
Similarly, the total deformation $y$ represents the current configuration $\mathcal{C}$, considering both plastic and elastic deformations.
Therefore, the elastic deformation describes the dislocation kinematics from $\mathcal{B}$ to $\mathcal{C}$.
Elastic deformation in mathematics is interpreted as an embedding of the Riemann--Cartan manifold $\mathcal{B}$ into Euclidean space $\R{3}$.
We derive the governing equation for the elastic embedding through the variational principle of hyperelastic materials.

According to the standard continuum mechanics, elastic deformation is typically expressed by the Green--Lagrange strain tensor $E$, which measures the geometrical difference between the current and reference configurations \cite{Marsden-Hughes}.
In the present framework, the strain tensor $E$ is defined as the difference of Riemannian metrics between the two configurations such that \cite{Marsden-Hughes,Grubic}
\begin{equation}\label{eq:strain}
E[y, \bi]=\frac{1}{2}\left(g[y]_{ij}-g[\bi]_{ij}\right)dx^i \otimes dx^j.
\end{equation}
Similarly, a local form of the elastic coefficient tensor $C$ is given by
\begin{equation}\label{eq:Cijkl}
C[\bi]=\left(\lambda\metric{\bi}^{ij}\metric{\bi}^{kl}
+\mu \metric{\bi}^{ik}\metric{\bi}^{jl}
+\mu \metric{\bi}^{il}\metric{\bi}^{jk}
\right)\pd{}{x^{i}}\otimes \pd{}{x^{j}} \otimes \pd{}{x^{k}} \otimes \pd{}{x^{l}},
\end{equation}
where $\lambda$ and $\mu$ are Lam\'e constants and $\metric{\bi}^{ij}$ represents the inverse matrix of Riemannian metric $\metric{\bi}_{ij}$.
For simplicity, we adopted the Cauchy solid condition $\lambda/\mu=1$.
The strain energy density of the St. Venant--Kirchhoff type hyperelastic material is defined as a quadratic form, given by $\mathcal{W}[\embc, \bi] = C[\bi](E[\embc, \bi], E[\embc, \bi])/2$.
By integrating the energy density over the manifold $\M$, we obtain the strain energy functional such that
\begin{align}\label{eq:StrainEnergyFunctional0}
  W[\embc] = \int_\M \frac{1}{2} C[\bi]^{ijkl} E[\embc,\bi]_{ij} E[\embc,\bi]_{kl} \upsilon[\bi].
\end{align}
Here, $\upsilon[\bi]$ denotes the volume form with respect to the intermediate configuration $\mathcal{B}$, defined by $\upsilon[\bi] = (\det\bi) dx^1 \wedge dx^2 \wedge dx^3$.
The hyperelastic material is isotropic and satisfies material frame indifference under nonlinear finite deformation.
The current configuration $\mathcal{C}$ is obtained from the total deformation $y$ that minimises the strain energy functional $W[y]$.

\subsection{Discretisation for numerical analysis}
\label{sec:iga}
The mechanics of dislocations can be formulated through two variational problems: the minimisation of plasticity and elasticity functionals, as described in equations (\ref{eq:plasticityfunctional0}) and (\ref{eq:StrainEnergyFunctional0}), respectively. 
These variational problems are computationally addressed using isogeometric analysis (IGA) \cite{hughes_iga}.
IGA employs non-uniform rational B-spline (NURBS) basis functions for variational problems.
As demonstrated in the subsequent section, the smoothness of the NURBS basis functions is essential in the current geometrical analysis.

\begin{figure}[ht]
    \centering
    \includegraphics[width=\textwidth]{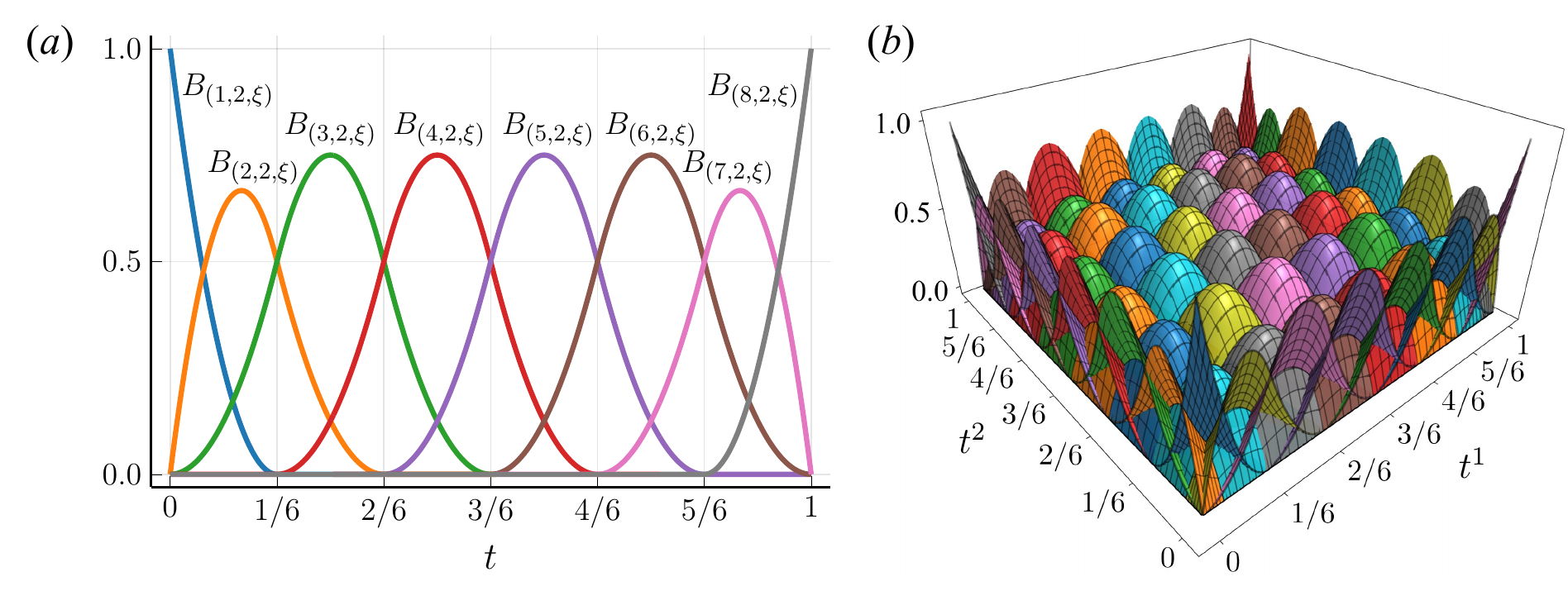}
    \caption{(a) 1-dimensional B-spline basis functions $B_{(i,2,\xi)}(t)$ of the second-order. The knot vector $\xi$ is defined by the non-decreasing sequence $\xi=(0,0,0,1/6,2/6,3/6,4/6,5/6,1,1,1)$. (b) 2-dimensional B-spline basis functions obtained by the product of $B_{(i,2,\xi)}(t)$ defined in (a). These basis functions satisfy the partitions of unity.}
    \label{fig:bsb}
\end{figure}

Let $p$ be a polynomial degree of the B-spline function and let $\xi=(\xi_1,\xi_2,\dots,\xi_m)$ be a non-decreasing sequence referred to as the knot vector.
The B-spline basis function is a set of $n=m-p-1$ piecewise polynomials $\{\bsb{i}{p}{\xi}\}_{i=1,\dots,n}$ defined on the interval $I_{\xi}=[\xi_1,\xi_m)$.
Here, the subscript $(i,p,\xi)$ denotes the element number $i$, polynomial degree $p$, and knot vector ${\xi}$, respectively.
Subsequently, the B-spline function is expressed by the Cox--de Boor recursion formula:
\begin{align}
\bsb{i}{p}{\xi}(t)=\frac{t-\xi_i}{\xi_{i+p}-\xi_{i}}\bsb{i}{p-1}{\xi}(t)+ \frac{\xi_{i+p+1}-t}{\xi_{i+p+1}-\xi_{i+1}}\bsb{i+1}{p-1}{\xi}(t),
\end{align}
for all $t\in[\xi_1,\xi_m)$.
The $0$-th order B-spline function $B_{(i,0,\xi)}(t)=1$ when $\xi_i \le t < \xi_{i+1}$, otherwise $B_{(i,0,\xi)}(t)=0$.
Figure \ref{fig:bsb} illustrates examples of B-spline basis functions.

NURBS basis functions are constructed from the rationalisation to $B_{(i,p,\xi)}(t)$.
Let $\hat{I}=I_{\xi^1}\times I_{\xi^2}\times I_{\xi^3}$ be the unit cube and let $\{B^\alpha(t)=\bsb{i_1}{p_1}{\xi_1}(t^1)\bsb{i_2}{p_2}{\xi_2}(t^2)\bsb{i_3}{p_3}{\xi_3}(t^3) \mid t=(t^1,t^2,t^3)\in \hat{I}\}$ be the three-dimensional B-spline basis functions.
Then the NURBS basis functions $N^{\alpha}(t)$ and NURBS map $x: \hat{I} \to \R{3}$
are defined by
\begin{align}\label{eq:nurbs}
  N^{\alpha}(t) = \frac{
    w^{\alpha}B^\alpha(t)
  }{
    \sum_{\beta=1}^{n} w^{\beta} B^\beta(t)
  },
  \quad
  x^i(t)=\sum_{\alpha=1}^n N^\alpha(t) a_\alpha^{i},
\end{align}
where the real-valued coefficients $\{w^\alpha\}$ are referred to as weights and each $(a_\alpha^1,a_\alpha^2,a_\alpha^3)\in \R{3}$ represents a control point that determines the NURBS map.

\section{Results of numerical analysis}
\subsection{Screw dislocation model and intermediate configuration}
\begin{figure}[htbp]
  \centering
  \includegraphics[width=\textwidth]{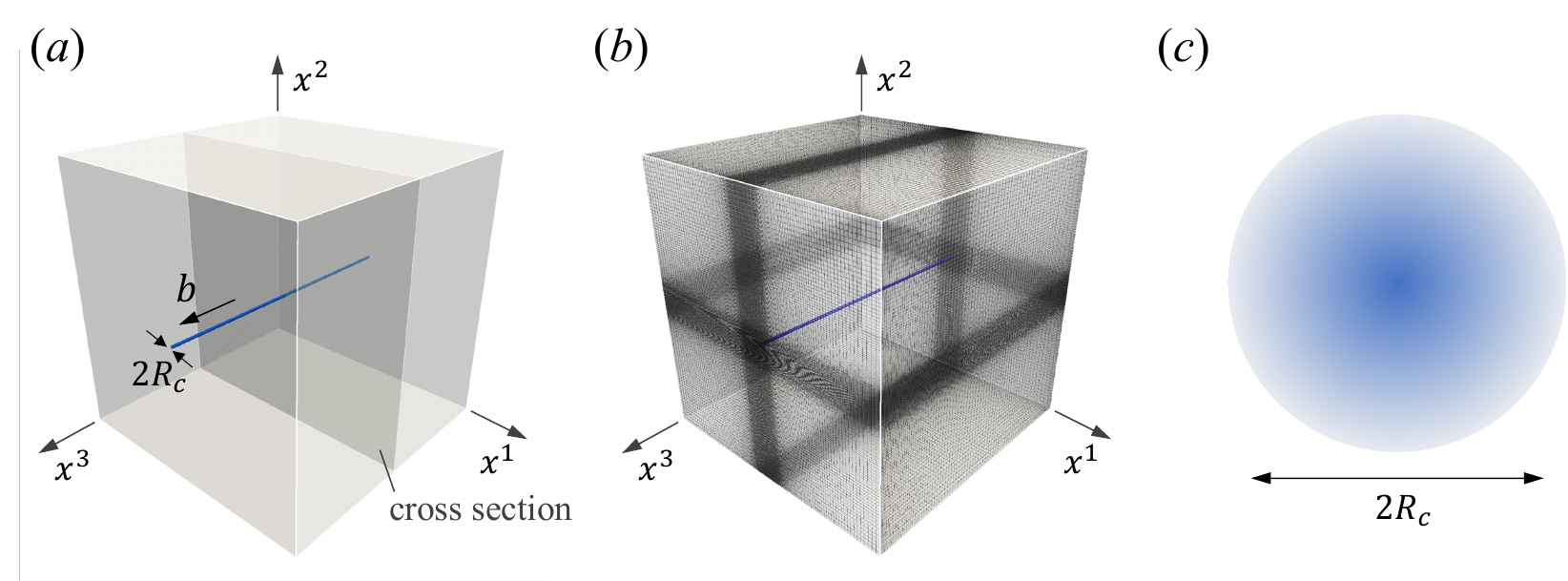}
  \caption{
        (a) Schematic illustration of the screw dislocation model.
        A straight dislocation line is positioned at the centre of the cubic model, with dimensions of $100 |b|\times 100 |b| \times 100 |b|$, oriented along the $x^3$-axis.
        (b) Non-uniform distribution of the NURBS basis function.
        To improve the precision of our calculations and capture the stress concentration near the dislocation core, we utilized a non-uniform basis function that concentrates in the vicinity of the dislocation line.
        (c) Cross-sectional distribution of dislocation density $\alpha$ within the dislocation core.
        The dislocation density exists only within the circular core with the radius $R_c$ and is absent outside.
    }
  \label{model}
\end{figure}

Figure \ref{model}(a) illustrates the model of screw dislocation used in this study.
The model is a cube with sides of length $100 |b|$, where $|b|$ represents the magnitude of the Burgers vector.
The dislocation line is positioned at the centre of the $x^1$--$x^2$ plane and is parallel to the $x^3$-axis.
The numerical analysis was performed using IGA, a Galerkin method that employs NURBS as basis functions.
The non-uniform distribution of knots for a total of $200\times 200\times 100$ basis is depicted in figure \ref{model}(b).
The B-spline has polynomial order of $p=2$, and the numerical analysis has over 12 million degrees of freedom.

Based on the data presented in figure \ref{model} (a), the Burgers vector $b$ and the tangent of dislocation line $n$ of the screw dislocation are $b^i=(0,0,b)$ and $n^i=(0,0,1)$, respectively.
Consequently, the dislocation density and corresponding torsion 2-form can be simplified to (see Equations (\ref{eq:dislocation_density}) and (\ref{eq:dislocation-torsion}))
\begin{align}\label{eq: dislocation densities}
    \alpha=f b dx^3 \otimes \rb{3}, \quad \tau=f b dx^1\wedge dx^2 \otimes \rb{3}.
\end{align}
The torsion for the screw dislocation has only one non-zero coefficient, $T^3_{12} = fb$, which is uniform along the $x^3$ direction.
The dislocation density distribution in the $x^1$--$x^2$ plane is determined by the radial distribution function $f$.
Let $r$ denote the distance from the dislocation centre within the plane.
We set the distribution function $f(r)$ such that 
\begin{align}\label{eq:radial_distribution}
    f(r)=\left\{
        \begin{array}{ll}
            \dfrac{3}{\pi R_c^2}\left( 1-\dfrac{r}{R_c} \right)&\quad(r \le R_c)\\
            0&\quad(r>R_c)
        \end{array}
    \right..
\end{align}
The function $f(r)$ has a cone-shaped distribution within the dislocation core of radius $R_c$ and is absent outside the core (see figure \ref{model} (c)).
The coefficient $3/\pi R_c^2$ was selected for normalisation.
In the limit of zero radius $R_c \to 0$, the current model converges to the classical Volterra dislocation theory.
We assume $R_c = |b| = 1$ unless otherwise stated to simplify the analysis.

\begin{figure}[htbp]
  \centering
  \includegraphics[width=1.0\textwidth]{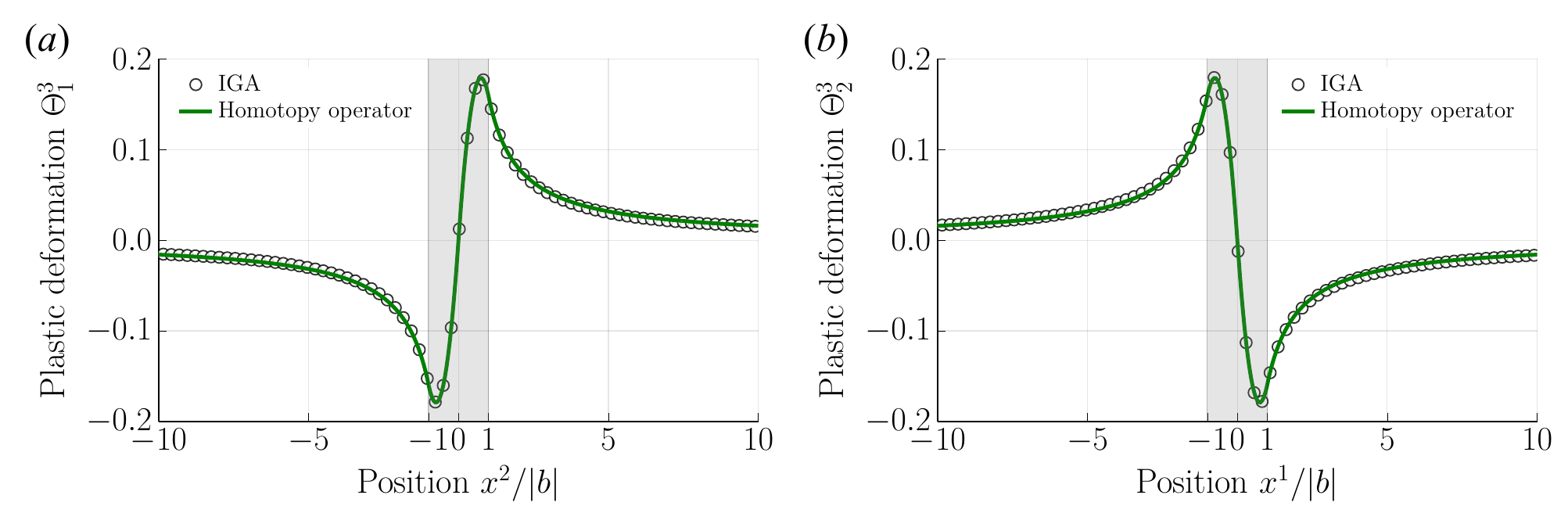}
  \caption{
Distribution of plastic deformation gradients (a) $\Theta^3_1$ and (b) $\Theta^3_2$ for the screw dislocation obtained from the central cross-section.
These plastic deformation gradients are continuously distributed throughout the entire region despite the source of the distribution being confined within the dislocation core. The current numerical analysis exhibits full concordance with analytical solutions obtained using the homotopy operator method.
}
  \label{fp line section}
\end{figure}

By substituting equation (\ref{eq: dislocation densities})$_2$ into (\ref{eq:plasticityfunctional0}) and numerically solving for the stationary point of the plasticity functional $I[\Theta]$, we derive the dual exact part of the plastic deformation gradient $\Theta$ attributable to the screw dislocation.
Figure \ref{fp line section} depicts the $\Theta^3_1$ and $\Theta^3_2$ distributions evaluated along the $x^1$ and $x^2$-axes on the central cross-section, as obtained from numerical analysis (see figure \ref{model} (a)).
The open circles represent the results of numerical analysis, while the solid curves depict the analytical solutions obtained using the homotopy operator (see \ref{AppendixA}\ref{sec:homotopy}).
Generally, the homotopy operator is applicable for the analytical integration of the Cartan first structure equation in an infinite medium \cite{edelen, acharya}.
Because the central cross-section of the current model remains unaffected by the surfaces due to the bilateral symmetry of the dislocation density, we can use the analytical solution for quantitative validation of the numerical results.
The numerical results align perfectly with the analytical solutions, validating the present numerical analysis.

\subsection{Plastic strain fields around screw dislocations}
By incorporating the numerical result $\Theta$ into equation (\ref{eq:Helmholtz decomposition}), we derive the bundle isomorphism $\bi$ (or plastic deformation gradient $F_p$), which defines the intermediate configuration $\mathcal{B}$.
First, we consider the Riemannian metric $g[\bi]$.
In continuum mechanics, the metric represents the right Cauchy--Green tensor of plastic deformation, \textit{i.e.} stress-free plastic strain due to the screw dislocation.

\begin{figure}[htbp]
  \centering
  \includegraphics[width=\textwidth]{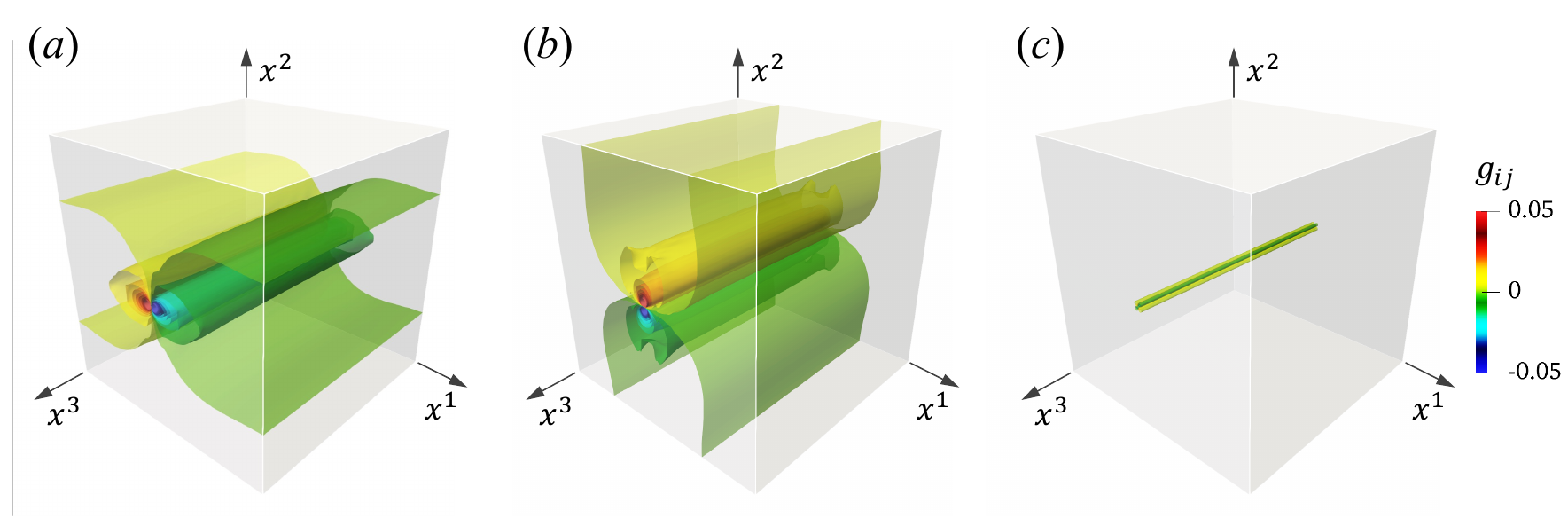}
  \caption{
    Distribution of Riemannian metric $\metric{\bi}$, or plastic strain, for the intermediate configuration around the screw dislocation.
    The figures depict the spatial distribution of
    (a) $\metric{\bi}_{23}=\metric{\bi}_{32}$,
    (b) $\metric{\bi}_{31}=\metric{\bi}_{13}$ and
    (c) $\metric{\bi}_{12}=\metric{\bi}_{21}$.
    Each contour plot includes 20 contour planes with equal division in the range $-5\times 10^{-2}< \metric{\bi}_{ij} <5\times 10^{-2}$.
    The strain components $\metric{\bi}_{23}$ and $\metric{\bi}_{31}$ exhibit a broad distribution throughout the domain, while $\metric{\bi}_{12}$ localises in the immediate vicinity along the dislocation line.
  }
  \label{fig: g[vartheta] isosurface}
\end{figure}

Figures \ref{fig: g[vartheta] isosurface} show the distribution of the off-diagonal components of $g[\bi]$.
These figures clearly illustrate that the dominant plastic strains, $g[\bi]_{23}$ and $g[\bi]_{31}$, exhibit high concentrations in the vicinity of the dislocation line.
However, their decay is gradual, and a significant portion of the strain extends to the free boundary.
This behaviour results from the boundary condition of the dual exact part: $\Theta |_{\partial \M}(\unitnormal{})=0$.
The plastic strains are distributed such that the iso-surfaces are perpendicular to the free surface.
This condition enables the transmission of plastic strain over long distances.
Free surfaces significantly affect the elastic deformation of continua.
However, the effect of free surfaces on plastic deformation has not yet been clarified, as far as the authors are aware.
By combining the Helmholtz decomposition and numerical calculations, the present study obtained the plastic deformation field with free boundaries for the first time.
In contrast, as shown in figure \ref{fig: g[vartheta] isosurface}(c), $g[\bi]_{12}$ exhibits a very different distribution from $g[\bi]_{23}$ and $g[\bi]_{31}$, and is only present in the immediate vicinity of the dislocation core.
The in-plane diagonal components $g[\bi]_{11}$ and $g[\bi]_{22}$ exhibit a similar trend.

The significant differences in the spatial distribution of plastic strain can be elucidated through the mathematical definition of the Riemannian metric.
According to the analytical solution obtained by the homotopy operator, the non-vanishing components of the dual exact part at the central cross-section are $\Theta^3_1$ and $\Theta^3_2$.
This indicates that these components are dominant in the entire domain.
If we substitute this into equations (\ref{eq:IntermediateMetric}) and (\ref{eq:Helmholtz decomposition}), we obtain 
\begin{align}
    \label{eq:screw g[bi]}
    g[\bi]_{ij}=\left(\delta_{ij} + \delta^3_i\Theta^3_j+\delta^3_j\Theta^3_i+\Theta^3_i\Theta^3_j\right)dx^i \otimes dx^j.
\end{align}
Hence, the coefficients $g[\bi]_{23}=\Theta^3_2$ and $g[\bi]_{13}=\Theta^3_1$ are linear, while $g[\bi]_{12}=\Theta^3_1 \Theta^3_2$ is quadratic concerning the dual exact part of the plastic deformation gradient $\Theta^i_j$.
The geometric nonlinearity of the right Cauchy--Green tensor results in $g[\bi]_{12}$.
Similar characteristics are confirmed in the diagonal components $g[\bi]_{11}$ and $g[\bi]_{22}$.
The remaining component becomes unity, $g[\bi]_{33}=1$, as predicted from equation (\ref{eq:screw g[bi]}).
Hence, the geometrical nonlinearity is responsible for the localisation of $g[\bi]_{12}$, $g[\bi]_{11}$ and $g[\bi]_{22}$ in the immediate vicinity of the core along the dislocation line.

\subsection{Elastic stress fields around screw dislocation}
The existence of torsion $T^3_{12}$ within the dislocation core explains the mathematical significance of the intermediate configuration $\mathcal{B}$.
This configuration is, however, a conceptual construct introduced through differential geometry, and the actual physical existence of a dislocation is in Euclidean space $\mathbb{R}^3$.
Therefore, we require the current configuration $\mathcal{C}$, which is obtained by embedding the intermediate configuration $\mathcal{B}$ into $\mathbb{R}^3$.
By minimising the elasticity functional in equation (\ref{eq:StrainEnergyFunctional0}) using the variational principle of hyperelastic materials, the embedding map $y$ can be determined.

\begin{figure}[htbp]
  \centering
  \includegraphics[width=0.95\textwidth]{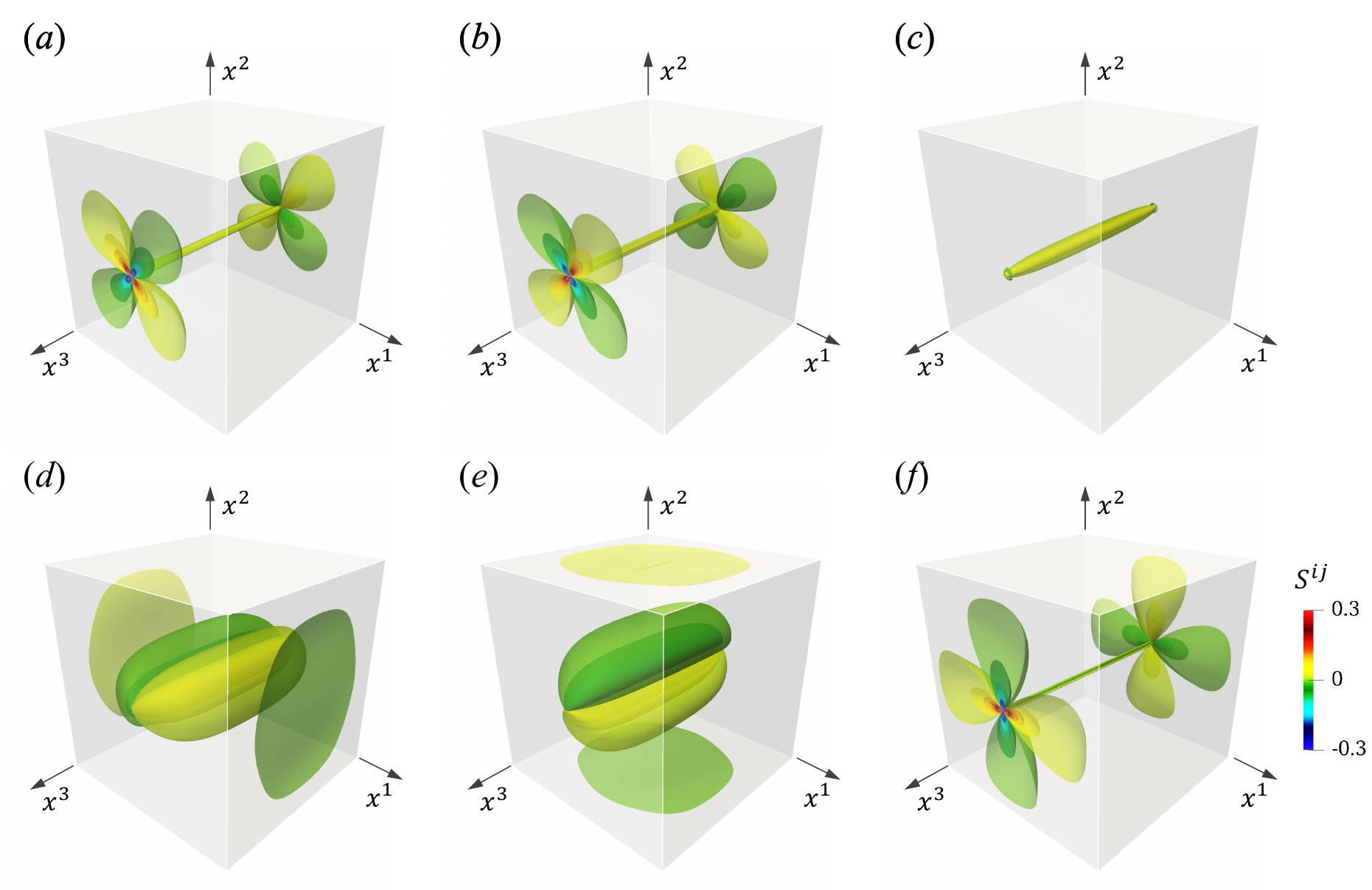}
  \caption{
    The second Piola--Kirchhoff stress fields $S^{ij}$ of the straight screw dislocation:
    (a) $S^{11}$,
    (b) $S^{22}$,
    (c) $S^{33}$,
    (d) $S^{23}=S^{32}$,
    (e) $S^{31}=S^{13}$ and
    (f) $S^{12}=S^{21}$.
    Stress fields are normalised by the scaling factor $D=\mu/2\pi$.
    Each contour plot includes 20 contour planes with equal divisions in the range $-0.3<S^{ij}/D<0.3$.
    The dominant stress components within the material are $S^{23}$ and $S^{31}$.
    The surface effect induces significant stress concentration at the dislocation line ends in (a), (b) and (d).
    It also confirms the distribution of weak normal stress $S^{33}$ in (c) the vicinity of the dislocation line.
  }
  \label{screw isosurface}
\end{figure}

Figures \ref{screw isosurface} illustrate the three-dimensional distribution of the second Piola--Kirchhoff stress $S^{ij}$ obtained from the screw dislocation.
The stress fields exhibit several intriguing properties absent in the classical Volterra dislocation.
For instance, the classical theory predicts that non-zero stress components are limited to $S^{23}\,(=S^{32})$ and $S^{31}\,(=S^{13})$. However, the present result reveals the generation of all stress components.
The dominant stress fields in the plane strain condition, \textit{i.e.} stress distribution on the central cross-section (see figure \ref{model} (a)), are $S^{23}$ and $S^{31}$.
However, in the vicinity of the free surfaces, these stresses rapidly diminish, and other stresses emerge.
The free surface significantly influences the stress fields $S^{ij}$.
Another noteworthy feature to discuss is the emergence of a localised stress field in the immediate vicinity of the dislocation core.
A prime example of this is the normal stress component, $S^{33}$, as depicted in figure \ref{screw isosurface}(c).
Similar trends are also observed in $S^{11}$, $S^{22}$, and $S^{12}$.
These stress components exhibit relatively small magnitudes and can be disregarded across the entire domain, except for the region directly beneath the surface.

\begin{figure}[htbp]
  \centering
  \includegraphics[width=1.0\textwidth]{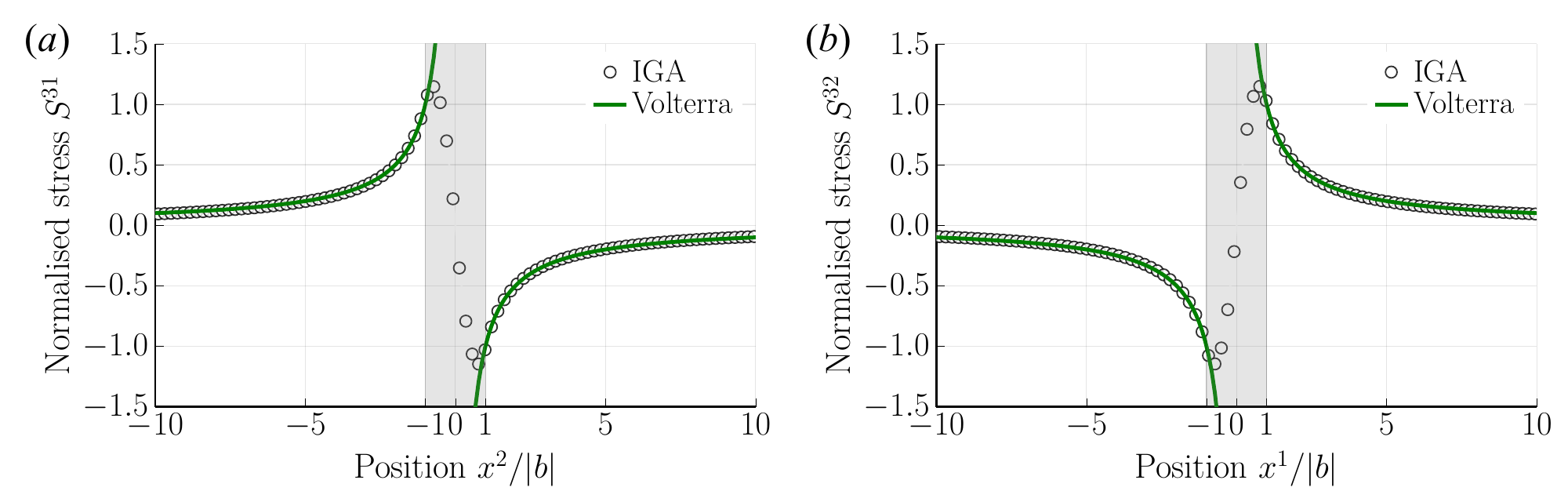}
  \caption{
    One-dimensional distribution of shear stresses, (a) $S^{31}$ and (b) $S^{32}$, obtained from the central cross-section.
    The open circles depict the current numerical results, while the solid curves represent the theoretical prediction based on the Volterra dislocation model.
    The results exhibit excellent quantitative agreement outside the dislocation core ($x^i/|b|>1$).
    In addition, the present result succeeded in removing the stress singularities at the dislocation centre $x^i/|b|=0$.
  }
  \label{fig:screw shear}
\end{figure}

Figure \ref{fig:screw shear} shows the cross-sectional distribution of the dominant shear stresses obtained at the central cross-section.
The open circles represent the numerical analysis results, while the solid curves depict the theoretical predictions given by the Volterra dislocation.
In conjunction with the dislocation configuration displayed in figure \ref{fp line section}, the central cross-section is subjected to the plane strain condition, enabling a direct quantitative comparison.
Excellent agreement between the two sets of results is observed outside the dislocation core ($x^i/b>1$), as depicted in the figures.
Notably, while the stresses predicted by the Volterra dislocation diverge to infinity within the dislocation core, the present results remain finite even at the dislocation centre.
The present geometrical theory has successfully eliminated the stress singularity by incorporating the continuous distribution of dislocation density $\alpha$.
The stress fields $S^{ij}$ are expected to approach the Volterra dislocation in the classical limit $R_c/|b| \to 0$, which is noteworthy.
It indicates that the core radius $R_c$ serves as a characteristic length scale for the dislocation.

\begin{figure}[htbp]
  \centering
  \includegraphics[width=0.95\textwidth]{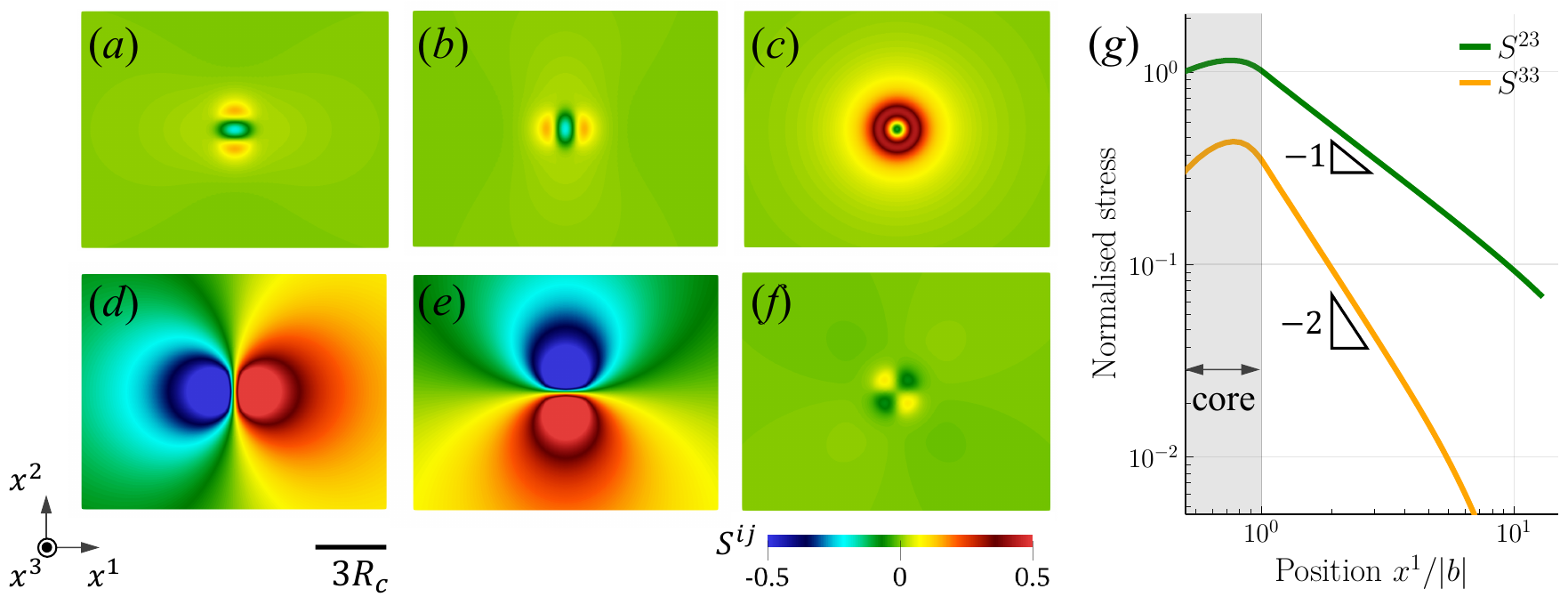}
  \caption{
Two-dimensional distribution of stress fields obtained from the central cross-section: (a) $S^{11}$, (b) $S^{22}$, (c) $S^{33}$, (d) $S^{23}$, (e) $S^{31}$, and (f) $S^{12}$.
All stress components are normalised by the scaling factor $D=\mu/2\pi$ such that $-0.5<S^{ij}/D<-0.5$.
Despite the classical Volterra dislocation model predicting the emergence of only two shear stresses, $S^{23}$ and $S^{31}$, the present results demonstrate the generation of all stress components due to geometrical nonlinearity.
(g) A double-logarithmic plot compares the stress decay rates for $S^{23}$ and $S^{33}$. While the decay rate of $S^{23}$ scales to $r^{-1}$, that of $S^{33}$ is $r^{-2}$.
  }
  \label{stress decay}
\end{figure}

The two-dimensional stress distributions obtained at the central cross-section are summarised in figures \ref{stress decay}(a)--(f).
The stress fields can be classified into three types of rotational symmetries: totally symmetric ($S^{33}$), two-fold symmetries ($S^{11}$, $S^{22}$, $S^{12}$), and two-fold symmetry with sign inversion ($S^{23}$, $S^{31}$).
The analysis demonstrates that the dominant shear stresses $S^{23}$ and $S^{31}$ are widely distributed, but other components decrease rapidly as the distance $r$ from the core increases. 
To quantitatively examine the differences in stress distribution, double-logarithmic plots of $S^{23}$ and $S^{33}$ are presented in figure \ref{stress decay}(g).
The results clearly demonstrate that the shear stress $S^{23}$ decays as $1/r$ outside the dislocation core, which is consistent with the classical Volterra dislocation.
However, the normal stress $S^{33}$ decays more rapidly, following a $1/r^2$ decay rate rather than $1/r$.
Similar behavior is confirmed in $S^{11}$, $S^{22}$, and $S^{12}$.
The elastic Green function, which is dependent on the elastic coefficients \cite{Mura} determines the decay rate of $1/r$ in linear elasticity.
Therefore, the stress decay rate of the Volterra dislocation is independent of the stress components.
However, as observed in equation (\ref{eq:Cijkl}) and figures \ref{fig: g[vartheta] isosurface}, present elastic coefficients incorporate the Riemannian metric $g[\bi]$, which exhibits a noticeable distribution around the dislocation line.
In the current geometrical framework, plastic strain influences the dislocation stress fields, which may explain the rapid decay of $1/r^2$ in $S^{33}$ and other stress components.

\section{Discussion}
\subsection{Geometrical frustration of dislocation}
The central theme of this study is the geometric origin of dislocation stress fields, which we explore in depth.
Our study of this subject is supported by modern terminology, specifically the concept of \textit{geometrical frustration} \cite{sadoc_mosseri}.
Geometrical frustration in a discrete system emerges when the local stable structure fails to achieve global stability due to geometric constraints.
Considering the continuum mechanics, plastic strain $g[\bi]$ results in incompatibility.
This outcome implies that the kinematics cannot be adequately expressed within the Euclidean space $\mathbb{R}^3$.
These insights form the geometric basis for introducing the Riemann--Cartan manifold.

The multiplicative decomposition of the total deformation gradient $F^i_j=\partial y^i/\partial x^j$ provides valuable insights into the role of geometrical frustration in the nonlinear mechanics of dislocation.
Consider $dx^i$ as the orthonormal dual basis on the reference configuration, and let $\bi^i=\bi^i_j dx^j=(F_p)^i_j dx^j$ and $dy^i=F^i_j dx^j$ be the linear transformations of $dx^i$ to the intermediate and current configurations.
The kinematics of dislocation can then be expressed through the multiplicative decomposition as follows:
\begin{align}\label{eq:multiplicative decomposition}
  dy^i
  =F^i_j dx^j
  =F^i_j \big[ (F_p^{-1})^j_k (F_p)^k_l \big] dx^l
  =F^i_j (F_p^{-1})^j_k \bi^k=(F_e)^i_j \bi^j,
\end{align}
where $F_p^{-1}$ represents the inverse of the plastic deformation gradient $F_p$, satisfying $(F_p^{-1})^i_k (F_p)^k_j=\delta^i_j$.
The last term in the equation describes the linear transformation of the dual basis upon the elastic embedding from the intermediate $\bi^i$ to the current configuration $dy^i$ through the coefficients $(F_e)^i_j=F^i_k (F_p^{-1})^k_j$.
Therefore, the coefficients $F_e$ define the elastic deformation gradient.
The decomposition reveals that the elastic deformation gradient is essential in resolving geometrical frustration.
As depicted in figure \ref{model} (c), the interior of the dislocation core cannot conform to the standard Euclidean space, while the Euclidean structure is maintained outside the core.
The inverse plastic deformation $F_p^{-1}$ in the elastic deformation is responsible for the elimination.
Our observations confirm that the geometrical frustration induced by $F_p^{-1}$ is precisely opposite to those resulting from $F_p$.
Hence, the inverse plastic deformation in $F_e$ effectively eliminates the geometrical frustration within the dislocation core while preserving the Euclidean geometry outside it.

The stress equilibrium equation gives rise to the second crucial role of $F_e$.
According to the variational principle of hyperelastic materials, elastic deformation must attain a state of minimal strain energy by satisfying the stress equilibrium equation.
However, the inverse plastic deformation gradient $F_p^{-1}$ itself might not be a general solution to the equation.
Therefore, the total deformation gradient $F^i_k$ is determined such that the composite map $F^i_k (F_p^{-1})^k_j$ satisfies the minimization of strain energy functional.
The total deformation gradient $F^i_k$ defines a map between the Euclidean manifolds of the reference and current configurations without introducing geometrical frustration.
This aspect explains the process by which the geometrical frustration yields elastic deformation, resulting in elastic strain and stress fields around the dislocation.

\subsection{Ricci curvature: direct origin of stress fields}
Our final objective is to address the symmetry of stress fields.
As elucidated in the preceding section, the dislocation stress fields $S^{ij}$ emerges from the geometrical frustration within the dislocation core.
Due to the geometrical equivalence of the torsion and dislocation density, the distribution of the frustration $T^3_{12}$ within the core is entirely symmetric (see figure \ref{model} (c)), implying no preferential orientation dependence.
Conversely, the stress fields $S^{ij}$ exhibit various rotational symmetries as depicted in figure \ref{stress decay}.
A mathematical explanation is necessary to explain the observed discrepancy in symmetry.

The mathematical properties of the Riemann--Cartan geometry can be fully utilised to provide a comprehensive explanation.
Recalling that we introduced the two mathematical representations for the intermediate configuration, Weitzenb\"ock manifold $(\mathcal{M}, g[\bi], \nabla_W[\bi])_{\mathcal{B}}$ and Riemannian manifold $(\mathcal{M}, g[\bi], \nabla_{LC}[\bi])_{\mathcal{B}}$.
Their mechanical states are indistinguishable because they share the Riemannian metric $g[\bi]$.
The mathematical expression for the geometrical frustration is the only point of discrepancy between the Weitzenb\"ock manifold and the Riemannian manifold. The former employs torsion (\ref{eq:torsion}) while the latter uses curvature (\ref{eq:curvature}).
Our screw dislocation is a Weitzenb\"ock manifold due to the inclusion of non-vanishing torsion $T^3_{12}$ within the dislocation core.
An intriguing aspect arises when characterising geometrical frustration using curvature instead of torsion.
This analysis necessitates the Levi--Civita connection $\nabla_{LC}$, which is defined by the differentiation to the Riemannian metric $g[\bi]$ (see equations (\ref{eq:IntermediateConnection}) and \ref{eq:ConnectionCoefficients})$_2$).
By utilising the NURBS basis function, we can feasibly differentiate and improve smoothness by increasing the polynomial order up to $p=4$.
The Riemannian manifold offers a significant advantage when examining the geometric origin of the stress field symmetry.

\begin{figure}[htbp]
  \centering
  \includegraphics[width=0.95\textwidth]{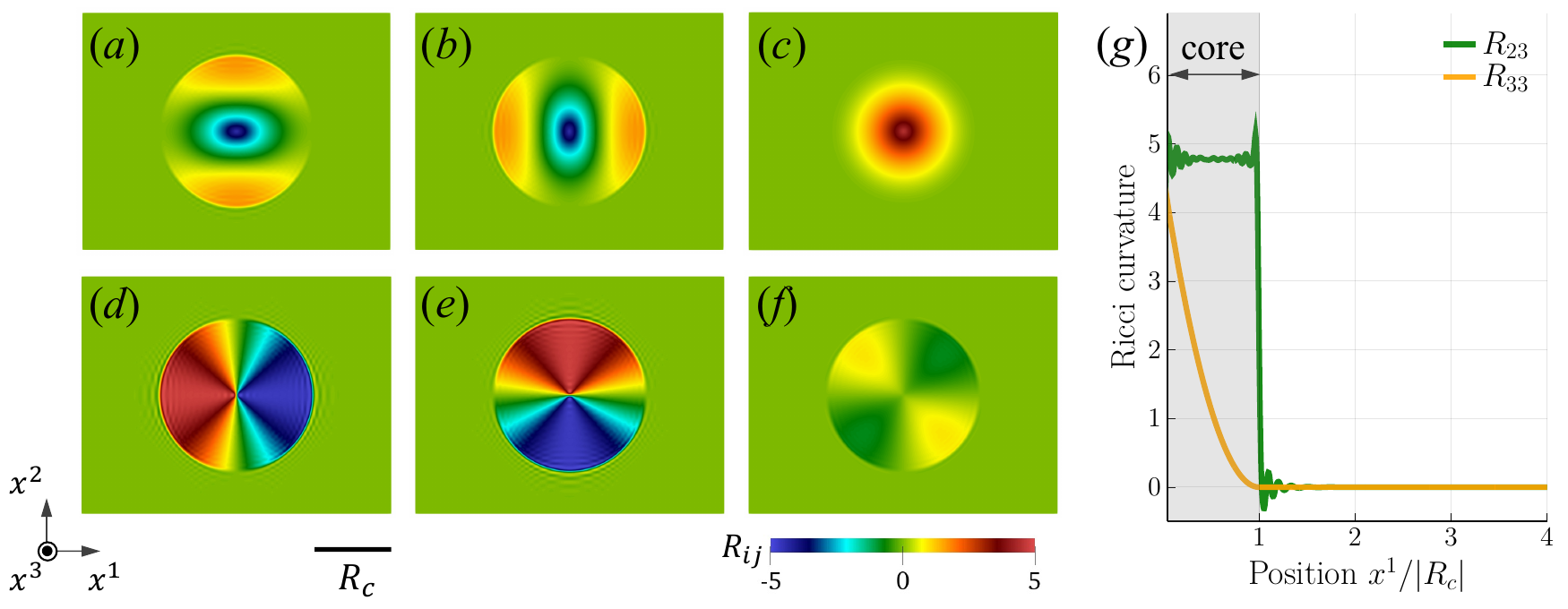}
  \caption{
Two-dimensional distribution of Ricci curvature $R_{ij}$ obtained from the central cross-section: (a) $R_{11}$, (b) $R_{22}$, (c) $R_{33}$, (d) $R_{23}$, (e) $R_{31}$, and (f) $R_{12}$.
Magnitude of $R_{11}$, $R_{22}$, $R_{33}$ and $R_{12}$ are magnified by factor 10 as they are weak compared to $R_{23}$ and $R_{31}$.
To evaluate the curvature tensor, we used the fourth-order B-spline basis functions as it requires higher derivatives (see equation (\ref{eq:ConnectionCoefficients})$_2$).
One-dimensional distributions of the curvatures $R_{23}$ and $R_{33}$ along the $x^1$-axis are shown in (g).
$R_{23}$ and $R_{33}$ are scaled by factor $-1$ and $10$, respectively.
Similar to the case of torsion, the curvatures are strictly confined within the dislocation core of radius $R_c$.
  }
  \label{plastic-torsion-curvature}
\end{figure}

Let $R^i_{jkl}$ be the coefficients of Riemannian curvature obtained from the differentiation of the metric $g[\bi]$.
While the kinematic information about geometrical frustration is included in the curvature, analysing all independent coefficients is not convenient.
Therefore, we introduce the Ricci curvature $\mathrm{Ric}[\bi]$, a natural by-product of $R^i_{jkl}$, as it fully encodes the geometrical information.
The local form of the Ricci curvature is defined by\cite{tu_differential_2017}
\begin{align}
\mathrm{Ric}[\bi]=R^k_{ikj}dx^i\otimes dx^j=R_{ij} dx^i\otimes dx^j,
\end{align}
where the coefficients $R_{ij}=R^k_{ikj}$ are the contraction of the Riemannian curvature.
The distribution of Ricci curvatures within the dislocation core is summarised in figures \ref{plastic-torsion-curvature} (a) to (g).
Similar to the case of torsion, the curvature coefficients only have non-zero values within the dislocation core.
This aspect indicates that geometrical frustration is confined solely to the interior of the dislocation core irrespective of their mathematical representations (torsion or curvature).
The coefficients of Ricci curvature $R_{ij}$ can be classified into three types of rotational symmetries: totally symmetric ($R_{33}$), two-fold symmetries ($R_{11}$, $R_{22}$, $R_{12}$), and two-fold symmetry with sign inversion ($R_{23}$, $R_{31}$).
The symmetries of $R_{ij}$ and stress $S^{ij}$ are identical (see figure \ref{stress decay}).
The coefficients with less symmetry, $R_{23}$ and $R_{31}$, exhibit significantly larger values than the remaining coefficients.
This aspect is also consistent with the magnitude of stresses.
Therefore, it is reasonable to conclude that the dislocation stress fields $S^{ij}$ result from the Ricci curvatures $R_{ij}$ within the dislocation core.
This circumstantial evidence is the positive proof of the long-lasting mathematical hypothesis: the duality between the Ricci curvature and internal stress tensor \cite{schaefer, minagawa, amari_durity}.

\section{Conclusion}

In this study, we performed theoretical modelling and numerical analysis of screw dislocation based on the differential geometry and the calculus of variations.
The distinctive feature of this theory is that it formulates dislocation kinematics using  Riemann--Cartan manifold.
The key findings are summarised as follows.

\begin{enumerate}
\item
In a dislocation theory based on differential geometry, kinematics can be expressed using the diffeomorphism of the Riemann--Cartan manifold, which is represented as a triplet $(\mathcal{M}, g[\cdot], \nabla[\cdot])$.
Here, the Riemannian metric $g[\cdot]$ and the affine connection $\nabla[\cdot]$ distinguish the three kinematic configurations: reference, intermediate, and current configurations.
The intermediate configuration, a state in which only plastic deformation occurred owing to dislocation and is excluded in the normal Euclidean space, is of particular importance.

\item
According to Kondo's theory, dislocation density in mechanics is equivalent to torsion in differential geometry.
By solving the Cartan first structure equation, we can construct the intermediate configuration for a given dislocation density distribution based on this observation.
The Helmholtz decomposition is essential in this process, breaking down the plastic deformation gradient into an exact part $dC^\infty(\M;\R{3})$ and a dual exact part $\mathcal{D}(\M;\R{3})$. This decomposition clarifies the boundary conditions for the intermediate configuration.
We express the Cartan first structure equation in variational form and solve it numerically using IGA.
Similarly, the elastic embedding from the intermediate to the current configuration is also solved using IGA by minimising the strain energy functional.
This framework enables the analysis of arbitrary configurations of dislocations.

\item 
The present analysis demonstrates the successful elimination of stress singularities at the centre of the screw dislocation.
The singularity-free stress fields exhibit remarkable quantitative agreement with the classical Volterra theory outside the dislocation core.
Additionally, we discovered normal stress fields attributed to geometrical nonlinearity.
The nonlinear stress fields are localised in the vicinity of the dislocation core, and the decay rate scales to $r^{-2}$, a rate significantly higher than the classical scaling rate of $r^{-1}$.

\item
The discovery is pivotal as it highlights the significance of geometrical frustration within the dislocation core. 
It denotes the extent to which the intermediate configuration deviates from Euclidean space, and elastic deformation is employed to restore the configuration to Euclidean space $\mathbb{R}^3$.
The elastic deformation gradient is introduced as a composition map involving the inverse plastic deformation gradient and the total deformation gradient through the multiplicative decomposition. The former eliminates geometrical frustration within the dislocation core, while the latter minimises elastic strain energy.
This analysis reveals that geometrical frustration is the direct cause of the dislocation stress fields.

\item
Leveraging the mathematical properties of the Riemann--Cartan manifold, the screw dislocation can be represented as a Riemannian manifold.
Consequently, the geometrical frustration within the dislocation core can be expressed by Ricci curvature rather than torsion.
The findings reveal that the six components of the Ricci curvature exhibit three distinct rotational symmetries, aligning perfectly with the symmetry of stresses.
This outcome unequivocally demonstrates that the dislocation stress fields are generated by the Ricci curvature within the dislocation core.
This result confirms the long-standing mathematical hypothesis of the stress--curvature duality.
\end{enumerate}

\vskip6pt

\appendix

\renewcommand{\thesection}{Appendix \Alph{section}}
\renewcommand{\theequation}{A.\arabic{equation}}
\section{}
\label{AppendixA}

\subsection{Metric and connection on the intermediate configuration}
\label{sec:A1}
This section explains how the bundle isomorphism $\bi$ induces the Riemannian metric $g[\bi]$ and affine connection $\nabla[\bi]$ on the intermediate configuration $\mathcal{B}$.
Suppose we have smooth vector fields $X, Y\in \sect{T\M}$ defined on the manifold $\mathcal{M}$.
Let $\bms{\bi}\colon \Gamma(T\M)\to \Gamma(\M\times \R{3})$ be a smooth map associated with $\bi$, and let $\bms{\bi}X$ be a smooth section of $\M\times \R{3}$.
More precisely, we have $\bms{\bi}{X}=\bms{\bi}(\spd{}{x^k})=\bi^i_k \rb{i}$, where $\rb{i}$ is the orthonormal basis on $\mathbb{R}^3$.
Then, the isomorphism induces the Riemannian metric $\metric{\bi}$ such that \cite{wenzelburger}
\begin{align}
\metric{\bi}\Pare{X, Y} \coloneqq \Angle{\bms{\bi}{X}, \bms{\bi}{Y}}_{\R{3}}.
\end{align}
This yields the Riemannian metric $g[\bi]$ shown in equation (\ref{eq:IntermediateMetric}).
Similarly, let $\M \times \mathbb{R}^3$ be a product bundle with a smooth section $s=s^i\rb{i}\in \Gamma(\M\times\R{3})$.
The product bundle equips a trivial connection $D: \Gamma(T\M)\times \Gamma(\M\times\R{3})\to \Gamma(\M\times\R{3})$ whose local form is given by $D_X s \coloneqq ds^i (X) \rb{i} = X^j \pd{s^i}{x^j}\rb{i}$.
Consequently, the bundle isomorphism $\bi$ induces the affine connection $\ac{\bi}$ such that
\begin{align}
    \ac{\bi}_X Y = \vartheta_\#^{-1} (D_{X} ({\bms{\bi}{Y}})).
\end{align}
A straightforward calculation confirms that the connection is compatible with the metric $g[\bi]$.
The flat, non-symmetric and $\metric{\bi}$-compatible affine connection $\ac{\bi}$ is referred to as the Weitzenb\"ock connection.
The local expression of the connection is shown in equation (\ref{eq:IntermediateConnection}).

\subsection{Helmholtz decomposition}
\label{sec:HelmholtzDecomposition}

Helmholtz decomposition of $\R{3}$-valued 1-forms $\df{1}{\M; \R{3}}$ plays a crucial role in constructing the variational formulation of the Cartan first structure equation.
It defines a direct sum decomposition of 1-forms as shown in equation (\ref{eq:Helmholtz decomposition}).
Here, formal definitions of the exact part $dC^\infty$ and the dual $\mathcal{D}$ are given by
\begin{align}\label{eq:AppendixHD1}
    dC^\infty(\mathcal{M};\R{3}) \coloneqq&{}
    \set{df\in\df{1}{\M;\R{3}}\mid \forall f\in C^\infty(\M;\R{3})},
    \\ \label{eq:AppendixHD2}
    \mathcal{D}(\mathcal{M};\R{3}) \coloneqq&{}
    \set{\omega\in\df{1}{\M;\R{3}}\mid \delta \omega=0, \omega(\mathcal{N})=0},
\end{align}
where $\mathcal{N}$ is the normal vector field on the boundary $\partial\M$.
The operator $\delta \coloneqq (-1)^{3(k+1)+1}*d*$ in equation (\ref{eq:AppendixHD2}) stands for the co-differential for $\R{3}$-valued $k$-forms $\df{1}{\M; \R{3}}$ and the symbol $*$ is Hodge star with respect to the Riemannian metric $g[x]$ on the reference configuration.
For instance, the Hodge star transforms $\mathbb{R}^3$-valued 1-form into 2-form in such a way that
\begin{align}\label{eq:hodge star for r3-valued 1-form}
  *\omega = *(\omega^i_jdx^j)E_i
  =(\omega^i_1 dx^2\wedge dx^3+\omega^i_2 dx^3\wedge dx^1+\omega^i_3 dx^1\wedge dx^2)E_i.
\end{align}
Similarly, the inner product of $\mathbb{R}^3$-valued 1-forms is given by $\Angle{\omega, \eta} = \delta_{ij} (\omega^i_1\eta^j_1 +\omega^i_2\eta^i_2 +\omega^i_3\eta^j_3)dx^1\wedge dx^2\wedge dx^3$.
This volume form defines the inner product of the manifold $\M$ such that
\begin{align}\label{eq:AppendixHD3}
  (\omega, \eta)\coloneqq \int_\M \Angle{\omega, \eta}.
\end{align}
Mathematical importance of the Helmholtz decomposition (\ref{eq:Helmholtz decomposition}) is that it is orthogonal with respect to the inner product (\ref{eq:AppendixHD3}) \cite{wenzelburger}.
It means that $\forall \psi \in dC^\infty(\mathcal{M};\R{3})$ and $\forall \Theta \in \mathcal{D}(\mathcal{M};\R{3})$, we have the orthogonality in the sense that $(\psi,\Theta)=0$.

\subsection{Homotopy operator for Cartan first structure equation}
\label{sec:homotopy}
We derive the analytical solution for the Cartan first structure equation for the screw dislocation in an infinite medium by using the homotopy operator \cite{edelen}.
Let $p_0$ and $p$ be points in $\R{3}$.
For $\lambda \in [0,1]$, we can define the homotopy $h_{p_0}(p, \lambda)=(1-\lambda)p_0+\lambda p$.
Intuitively, $h_{p_0}$ is the straight line in $\R{3}$ parametrised by $\lambda$ with start and end points $p_0$ and $p$.
Let $S_{p_0}\subset \R{3}$ be a sub-space of $\R{3}$ that contains $p_0$.
Then, $S_{p_0}$ is called star-shaped region centred at $p_0$ if, for arbitrary $p\in S_{p_0}$, the line $h_{p_0}(p, \lambda)$ is the subset of $S_{p_0}$.
The homotopy operator centred at $p_0$ is a map $H_{p_0}\colon \df{k}{S_{p_0}}\to \df{k-1}{S_{p_0}}$ defined as 
\begin{align}
    \forall \omega\in \df{k}{S_{p_0}},\quad H_{p_0}\omega = \Angle{X, \int_0^1 h^*_{p_0,\lambda}\Pare{\frac{\omega}{\lambda}}d\lambda},
\end{align}
where $X=(x^i-p_0^i)\pd{}{x^i}$ is the vector field on $S_{p_0}$, $h^*_{p_0,\lambda}$ is the pull-back and $\Angle{\cdot, \cdot}$ is the product of a vector field and a differential form.
The homotopy operator satisfies the relation $dH_{p_0}+H_{p_0}d=\mathrm{id}_{\df{k}{S_{p_0}}}$ for $k\geq 1$, where $d$ is the exterior derivative and $\mathrm{id}_{\df{k}{S_{p_0}}}$ is the identity map in $\df{k}{S_{p_0}}$.
 
We apply the homotopy operator to the Cartan first structure equation for the bundle isomorphism $\bi{}$.
Since $\bi^i$ is the 1-form on $\M$, it satisfies $\bi{}^i=dH_{p_0}\bi{}^i+H_{p_0}d\bi{}^i=dH_{p_0}\bi{}^i+H_{p_0}\tau^i$.
By definition, the term $H_{p_0}\bi{}^i$ is the smooth function on $S_{p_0}$ and it does not contribute to the torsion since $ddH_{p_0}\bi{}^i=0$.
Therefore, we set $x^i=H_{p_0}\bi^i$ so that the intermediate configuration coincides with the reference configuration when $\tau^i=0$.
As a consequence, we obtain the analytical solution of the Cartan first structure equation such that $\bi{}^i=dx^i+H_{p_0}\tau^i$.

In the case of present screw dislocation, we can derive the analytical expression of $\bi{}^i$.
Here we take the centre of the star-shaped region as $p_0=(0,0, x^3)$, \textit{i.e.} $p_0$ is located on the dislocation line.
Straightforward calculation reveals that the pullback reads $h^*_{p_0,\lambda}\tau^3 = bf(\lambda r) \lambda^2 dx^1\wedge dx^2$.
Hence, we obtain
\begin{align}
    H_{p_0}\tau^3 =&{} \Angle{X, b\int_0^1 f(\lambda r) \lambda
    d\lambda}dx^1\wedge dx^2 = \frac{b\Psi}{2\pi}(-x^2dx^1 +x^1 dx^2),
\end{align}
where $\Psi = \int_0^1f(\lambda r) \lambda d\lambda$ is a function calculated as follows:
\begin{align}
    \Psi = \begin{cases}
        \dfrac{3R_c-2r}{R_c^3} \quad &(r\leq R_c)\\
        \dfrac{1}{ r^2} \quad &(r>R_c)
    \end{cases}.
\end{align}
We obtain the non-zero components analytically such that $\Theta^3_1=-\frac{b\Psi}{2\pi}x^2$ and $\Theta^3_2=\frac{b\Psi}{2\pi}x^1$ by comparing the coefficients with the Helmholtz decomposition $\bi^i=dx^i+\Theta^i_j dx^j$ (see \cite{kobayashi_1, kobayashi_2} for details).
These expressions satisfy $\delta \Theta=0$ and $d\Theta=\tau$ as required.

\enlargethispage{20pt}

%\ethics{Insert ethics text here.}

\providecommand{\dataccess}[1]
{
  \small	
  \textbf{{Data Accessibility.}} #1
}
\providecommand{\aucontribute}[1]
{
  \small	
  \textbf{{Authors' Contributions.}} #1
}
\providecommand{\competing}[1]
{
  \small	
  \textbf{{Competing Interests.}} #1
}
\providecommand{\funding}[1]
{
  \small	
  \textbf{{Funding.}} #1
}
\providecommand{\ack}[1]
{
  \small	
  \textbf{{Acknowledgements.}} #1
}

\noindent
\funding{This work was supported by JST, PRESTO Grant Number JPMJPR1997 Japan and JSPS KAKENHI Grant Numbers JP18H05481, JP23K13221.}

\noindent
\ack{
  The authors gratefully acknowledge Dr A. Suzuki for discussions on numerical analysis.
  }


\begin{thebibliography}{9}
\bibitem{Hirth} Anderson PM, Hirth JP, Lothe J. 2017 Theory of dislocations. 2017 edn. Cambridge: Cambridge University Press. 
\bibitem{Volterra} Volterra V. 1907 Sur l'\'equilibre des corps \'elastiques multiplement connexes. Ann. Sci. \'Ecole Norm. Sup. 24, 401--517. (doi:10.24033/asens.583)
\bibitem{orowan3} Orowan E. 1934 Zur Kristallplastizit\"at. III. Zeitschrift f\"ur Physik 89, 634--659. (doi:10.1007/BF01341480)
\bibitem{taylor} Taylor GI. 1934 The mechanism of plastic deformation of crystals. Part I.---Theoretical. Proc. R. Soc. Lond. A 145, 362--387. (doi:10.1098/rspa.1934.0106)
\bibitem{polanyi} Polanyi M. 1934 \"Uber eine Art Gitterst\"orung, die einen Kristall plastisch machen k\"onnte. Zeitschrift f\"ur Physik 89, 660--664. (doi:10.1007/BF01341481)
\bibitem{Mura} Mura T., Micromechanics of defects in solids. 1998 Dordrecht: Kluwer Academic Publishers.
\bibitem{lazar_1} Lazar M. 2003 A nonsingular solution of the edge dislocation in the gauge theory of dislocations. Journal of Physics A: Mathematical and General 36, 1415--1437. (doi:10.1088/0305-4470/36/5/316)
\bibitem{lazar_2} Lazar M, Maugin GA. 2005 Nonsingular stress and strain fields of dislocations and disclinations in first strain gradient elasticity. International Journal of Engineering Science 43, 1157--1184. (doi:10.1016/j.ijengsci.2005.01.006)
\bibitem{lazar_3} Lazar M, Maugin GA, Aifantis EC. 2006 Dislocations in second strain gradient elasticity. International Journal of Solids and Structures 43, 1787--1817. (doi:10.1016/j.ijsolstr.2005.07.005)
\bibitem{lazar_4} Lazar M. 2017 Non-singular dislocation continuum theories: strain gradient elasticity vs. Peierls--Nabarro model. Philosophical Magazine 97, 3246--3275. (doi:10.1080/14786435.2017.1375608)
\bibitem{Marsden-Hughes} Marsden JE, Hughes TJR, Carlson DE. 1984 Mathematical Foundations of Elasticity. New York: Dover publications, inc. (doi:10.1115/1.3167757)
\bibitem{kondo_1} Kondo K. 1955 Geometry of elastic deformation and incompatibility. Gakujutsu Bunken Fukyukai.
\bibitem{kondo_2} Kondo K. 1955 Non-Riemannian geometry of imperfect crystals from a macroscopic viewpoint. Gakujutsu Bunken Fukyukai. 
\bibitem{bilby_continuous_1956} Bilby BA, Smith E. 1956 Continuous Distributions of Dislocations. III. Proceedings of the Royal Society of London. Series A, Mathematical and Physical 236, 26.
\bibitem{bilby_notitle_1955} Bilby BA, Bullough R, Smith E. 1955 Continuous distributions of dislocations: a new application of the methods of non-Riemannian geometry. Proceedings of the Royal Society of London. Series A. Mathematical and Physical Sciences 231, 263--273. (doi:10.1098/rspa.1955.0171)
\bibitem{kroner_nicbt-lineare_nodate} Kr\"oner E, Seeger A. 1959 Nicht-lineare Elastizit\"atstheorie der Versetzungen und Eigenspannungen., 23.
\bibitem{amari} Amari S. 1962 On Some Primary Structures of Non-Riemannian Plasticity Theory. RAAG Memoirs 3, 163--172.

\bibitem{anthony_theorie_1970} Anthony K-H. 1970 Die theorie der disklinationen. Arch. Rational Mech. Anal. 39, 43--88. (doi:10.1007/BF00281418)


\bibitem{noll} Noll W. 1967 Materially uniform simple bodies with inhomogeneities. Archive for Rational Mechanics and Analysis 27, 32.
\bibitem{wang} Wang CC. 1967 On the geometric structures of simple bodies, a mathematical foundation for the theory of continuous distributions of dislocations. Archive for Rational Mechanics and Analysis 27, 62.
\bibitem{dewit} de Wit R. 1981 A view of the relation between the continuum theory of lattice defects and non-euclidean geometry in the linear approximation. International Journal of Engineering Science 19, 1475--1506. (doi:10.1016/0020-7225(81)90073-2)
\bibitem{le_stumpf_1} Le KCh, Stumpf H. 1993 Constitutive equations for elastoplastic bodies at finite strain: thermodynamic implementation. Acta Mechanica 100, 155--170. (doi:10.1007/BF01174787)
\bibitem{le_stumpf_2} Le KC, Stumpf H. 1996 A model of elastoplastic bodies with continuously distributed dislocations. International Journal of Plasticity 12, 611--627. (doi:10.1016/S0749-6419(96)00022-8)

\bibitem{wenzelburger} Wenzelburger J. 1998 A kinematic model for continuous distributions of dislocations. Journal of Geometry and Physics 24, 334--352. (doi:10.1016/S0393-0440(97)00016-8)
\bibitem{binz} Binz E, Schwarz G, Wenzelburger J. 2001 On the dynamics of continuous distributions of dislocations. Quart. Appl. Math. 59, 225--239. (doi:10.1090/qam/1827812)


\bibitem{yavari_riemanncartan_2012} Yavari A, Goriely A. 2012 Riemann--Cartan geometry of nonlinear dislocation mechanics. Archive for Rational Mechanics and Analysis 205, 59--118. (doi:10.1007/s00205-012-0500-0)
\bibitem{yavari_riemanncartan_2013} Yavari A, Goriely A. 2013 Riemann--Cartan geometry of nonlinear disclination mechanics. Mathematics and Mechanics of Solids 18, 91--102. (doi:10.1177/1081286511436137)
\bibitem{yavari_geometry_2014} Yavari A, Goriely A. 2014 The geometry of discombinations and its applications to semi-inverse problems in anelasticity. Proceedings of the Royal Society A: Mathematical, Physical and Engineering Sciences 470, 20140403--20140403. (doi:10.1098/rspa.2014.0403)

\bibitem{edelen} Edelen D, Lagoudas D. 1988 Gauge theory and defects in solids. Burlington, MA: Elsevier.

\bibitem{acharya} Acharya A. 2001 A model of crystal plasticity based on the theory of continuously distributed dislocations. Journal of the Mechanics and Physics of Solids 49, 761--784. (doi:10.1016/S0022-5096(00)00060-0)
\bibitem{clayton} Clayton JD. 2015 Defects in nonlinear elastic crystals: differential geometry, finite kinematics, and second-order analytical solutions. ZAMM - Journal of Applied Mathematics and Mechanics / Zeitschrift f\"ur Angewandte Mathematik und Mechanik 95, 476--510. (doi:10.1002/zamm.201300142)

\bibitem{tu_differential_2017} Tu LW. 2017 Differential geometry: connections, curvature, and characteristic classes. New York, NY: Springer Science+Business Media. 
\bibitem{nakahara_geometry_2003} Nakahara M. 2003 Geometry, Topology and Physics, Second Edition. Taylor \& Francis. (doi:10.1201/9781420056945)
\bibitem{nye} Nye JF. 1953 Some geometrical relations in dislocated crystals. Acta Metallurgica 1, 10.

\bibitem{schwarz} Schwarz G. 1995 Hodge decomposition---a method for solving boundary value problems. Berlin, Heidelberg: Springer Berlin Heidelberg. (doi:10.1007/BFb0095978)

\bibitem{Grubic} Grubic N, LeFloch PG, Mardare C. 2014 The equations of elastostatics in a Riemannian manifold. Journal de Math\'ematiques Pures et Appliqu\'ees 102, 1121--1163. (doi:10.1016/j.matpur.2014.07.009)

\bibitem{hughes_iga} Hughes TJR, Cottrell JA, Bazilevs Y. 2005 Isogeometric analysis: CAD, finite elements, NURBS, exact geometry and mesh refinement. Computer Methods in Applied Mechanics and Engineering 194, 4135--4195. (doi:10.1016/j.cma.2004.10.008)

\bibitem{sadoc_mosseri} Sadoc J-F, Mosseri R. 1999 Geometrical Frustration. Cambridge University Press. (doi:10.1017/CBO9780511599934)

\bibitem{schaefer} Schaefer H. 1953 Die Spannungsfunktionen des dreidimensionalen Kontinuums und des elastischen K\"orpers. Z. angew. Math. Mech. 33, 356--362. (doi:10.1002/zamm.19530331006)
\bibitem{minagawa} Minagawa S. 1962 Riemannian three-dimensioal stress-function space. RAAG Memoirs 3, 69--81.
\bibitem{amari_durity} Amari S, Kou K. 1963 Dual Dislocations and Non-Riemmannian Stress Space. RAAG Research Notes, 3rd Ser. 82.

\bibitem{kobayashi_1} Kobayashi S, Tarumi R. 2021 Geometrical modeling and numerical analysis of screw dislocation. Transactions of the JSME (in Japanese) 87, 20-00409-20--00409. (doi:10.1299/transjsme.20-00409)

\bibitem{kobayashi_2} Kobayashi S, Tarumi R. 2021 Geometrical modeling and numerical analysis of edge dislocation. Transactions of the JSME (in Japanese) 87, 21-00031-21--00031. (doi:10.1299/transjsme.21-00031)


\end{thebibliography}
\end{document}